\documentclass[journal]{IEEEtran}

\usepackage[font=footnotesize]{subfig}
\usepackage{amssymb}
\usepackage{amsfonts}
\usepackage{tikz}
\usepackage{cite}
\usepackage[cmex10]{amsmath}
\usepackage{url}
\usepackage{multirow}
\usepackage{ifthen}
\usepackage{arydshln}

\begin{document}

\title{New Combinatorial Construction Techniques\\for Low-Density Parity-Check Codes and\\Systematic Repeat-Accumulate Codes}

\author{Alexander~Gruner and Michael~Huber,~\IEEEmembership{Member,~IEEE} %
\thanks{Manuscript received July 15, 2011; revised March 28, 2012.
The work of A.~Gruner was supported within the Promotionsverbund `Kombinatorische Strukturen' (LGFG Baden-W\"urttemberg).
The work of M.~Huber was supported by the Deutsche Forschungsgemeinschaft (DFG) via a Heisenberg grant (Hu954/4) and a Heinz Maier-Leibnitz Prize grant (Hu954/5).}
\thanks{The authors are with the Wilhelm Schickard Institute for Computer Science, Eberhard Karls Universit\"at T\"ubingen, Sand~13,
D-72076 T\"ubingen, Germany (corresponding author's e-mail address: michael.huber@uni-tuebingen.de).}}%


\maketitle

\newtheorem{theorem}{Theorem}
\newtheorem{example}{Example}

\begin{abstract}
This paper presents several new construction techniques for low-density parity-check (LDPC) and systematic repeat-accumulate (RA) codes. Based on specific classes of combinatorial designs, the improved code design focuses on high-rate structured codes with constant column weights 3 and higher. The proposed codes are efficiently encodable and exhibit good structural properties. Experimental results on decoding performance with the sum-product algorithm show that the novel codes offer substantial practical application potential, for instance, in high-speed applications in magnetic recording and optical communications channels.
\end{abstract}


\section{Introduction}\label{intro}

\IEEEPARstart{S}{tructured} low-density parity-check (LDPC) codes have been designed recently based on finite geometries and, more generally, balanced incomplete block designs (BIBDs), see, for example,~\cite{Kou01,Hon04,Vas04, JohnDiss, Well03}. One of the main advantages of these structured LDPC codes is that they can lend themselves to very low-complexity encoding, as opposed to random-like LDPC codes. Moreover, experimental results show that the proposed codes perform well with iterative decoding.

In the first part of this paper, we present novel high-rate structured LDPC codes with constant column weights~3 and higher, based on cyclic BIBDs (CBIBDs), resolvable BIBDs (RBIBDs), and cyclically resolvable cyclic BIBDs (CRCBIBDs). We obtain several infinite classes of $(k,r)$-regular LDPC codes with values of $k$ varying from 3 to 8 (Theorems~\ref{thm_CDF-LDPC Codes},~\ref{thm_RBIBD-LDPC Codes}, and~\ref{thm_CRCBIBD-LDPC Codes}) as well as one infinite class of $(k,r)$-regular LDPC codes for $k$ any prime power (Theorem~\ref{thm_RBIBD-LDPC Codes_prime}), all admitting flexible choices of $r$ and the code length. The presented results are more general than previous ones \mbox{in~\cite{Hon04,Vas04,JohnDiss,Well03}}.
Our proposed LDPC codes have good structural properties, their Tanner graphs are free of short cycles of length 4 and have girth 6.
Their code rate is high, achieving $9/10$ or higher already at small block lengths. This is typically the rate of interest in high-speed applications, for example, in magnetic recording and optical communications channels. Most classes of the constructed codes have a quasi-cyclic (or similar) structure, which allows linear-complexity encoding with simple feedback shift registers. In addition, many classes exhibit very sparse parity-check matrices. Experimental results on decoding performance show that the novel LDPC codes perform very well with the sum-product algorithm~\cite{mncN},
significantly better than known BIBD-LDPC codes and random Gallager LDPC codes~\cite{gal62} for short to moderate block lengths (cf.~Fig.~\ref{table_Simu1}). We observe furthermore, very interestingly, for the proposed codes a performance gain as the column weights grow larger, and a particular good performance of LDPC codes based on CRCBIBDs.

\begin{figure*}[!t]
	\centerline{
		\subfloat{
			\includegraphics[scale = 0.8]{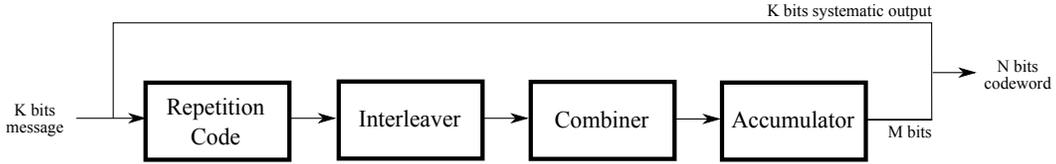}
		}
	}
	\caption{Low-complexity encoder of systematic RA codes}
	\label{encoder}
\end{figure*}

In the second part of the paper, we apply our combinatorial construction techniques to systematic repeat-accumulate (RA) codes. RA codes were first introduced in~\cite{turbo_like_codes} as a serial concatenation of a rate-$\frac{1}{q}$ repetition code and a \mbox{rate-1} convolutional code with transfer function $\frac{1}{1+D}$, called the accumulator (Fig. \ref{encoder}). Between these two constituent codes, an interleaver permutes the output of the repetition code and in some cases, the encoding scheme is also refined by a rate-$a$ combiner. This combiner performs modulo-2 addition on sets of $a$ bits. Here, we consider systematic RA codes (abbreviated from now on as \emph{sRA codes}), where the message of length $K$ is concatenated with the output of the accumulator of length $M$. For a detailed description of the encoding process, we refer the reader to~\cite{pract_interleavers_johnson_weller}. The scheme leads to a parity-check matrix of the form $H = [H_1 H_2]$, where $H_1$ is
  an $M\times K$ matrix with column weight $q$ and row weight $a$, and $H_2$ is an $M\times M$ double diagonal matrix designated solely by the accumulator~\cite{pract_interleavers_johnson_weller}. It is obvious that $H$ is sparse and thus the code can be decoded with the sum-product algorithm in the same way as for LDPC codes.

One drawback of sRA codes is that they possess many weight-2 columns in $H_2$, which has a negative effect on the minimal distance of the code and thus on the decoding performance. These low-weight columns result solely from the accumulator structure and hence Johnson and Weller~\cite{combinatorial_interleavers_for_systematic_ra_codes} have introduced a new accumulator type to increase the weights of these columns. The new accumulator produces many columns of weight 3 instead of weight 2 and, consequently, the resulting codes are called \mbox{weight-3 RA (w3RA)} codes. In the present paper, we propose a generalized accumulator structure, leading to novel coding schemes, termed  \emph{weight-$q$ RA (wqRA) codes}. This allows code constructions for higher column weights, having the advantage that their parity-check matrices are even more similar to those of regular LDPC codes and thus have a comparable gain in decoding performance (cf.~Fig.~\ref{table_Simu2}). We note that, while BIBDs by their axioms can always be used to build up parity-check matrices of sRA codes, it is the specific nature of CBIBDs, RBIBDs and CRCBIBDs that is exploited to build up parity-check matrices of wqRA codes.

The paper is organized as follows: Section~\ref{LDPC_BIBD} briefly introduces BIBDs and their usefulness in general LDPC code design. We give furthermore a short overview of previous results on LDPC code design based on CBIBDs, RBIBDs and CRCBIBDs. In Section~\ref{new}, we extend these previous results and present several novel high-rate LDPC codes with higher column weights. Experimental results on decoding performance are given in Subsection~\ref{sim}. 
In Section~\ref{new2}, we propose a generalized accumulator structure for sRA codes, leading to our novel coding schemes of wqRA codes. We develop several code constructions from CBIBDs, RBIBDs and CRCBIBDs that are capable of producing such wqRA codes. Simulation results are presented in Subsection~\ref{sim2}. The paper is concluded in Section~\ref{concl}.

\section{General LDPC Code Design Based on Balanced Incomplete Block Designs}\label{LDPC_BIBD}

\subsection{Balanced Incomplete Block Designs}\label{tools}

We give some standard material on combinatorial designs that is important for our further purposes. For encyclopedic references, we refer the reader to~\cite{BJL1999,crc06}. Various applications of combinatorial designs to coding, cryptography and information security are described in~\cite{Hu2009,Hu2010}, including many further references.

Let $X$ be a set of $v$ elements and $\mathcal{B}$ a collection of \mbox{$k$-subsets} of $X$. The elements of $X$ and $\mathcal{B}$ are called \emph{points} and \emph{blocks}, respectively. An ordered pair \mbox{$(X,\mathcal{B})$} is defined to be a \emph{balanced incomplete block design} (or \emph{2-design}), and denoted by $\mbox{BIBD}(v,k,\lambda)$, if each pair of points is contained in exactly $\lambda$ blocks. A $\mbox{BIBD}(v,k,1)$ is also called a \emph{Steiner} 2-design (or system). It is straightforward that in a BIBD each point is contained in the same number $r$ of blocks, and for the total number $b$ of blocks, the parameters of a BIBD satisfy the relations $bk=vr$ and $\lambda(v-1)=r(k-1)$.
For $(X,\mathcal{B})$ a $\mbox{BIBD}(v,k,\lambda)$, its \emph{incidence matrix} is a  $v \times b$ matrix
$A=(a_{ij})$ ($1\leq i \leq v$, $1\leq j \leq b$), in which $a_{ij}=1$ if the $i$-th point of $X$ is contained in the $j$-th block of $\mathcal{B}$, and $a_{ij}=0$ otherwise. Clearly, $A$ is unique up to column and row permutation.

Let $(X,\mathcal{B})$ be a $\mbox{BIBD}(v,k,\lambda)$, and let $\sigma$ be a permutation on $X$. For a block $B=\{b_1,\ldots,b_k\} \in \mathcal{B}$, define $B^\sigma :=\{b^\sigma_1,\ldots,b^\sigma_k\}$. If $\mathcal{B}^\sigma:=\{B^\sigma : B \in \mathcal{B}\}=\mathcal{B}$, then $\sigma$ is called an \emph{automorphism} of $(X,\mathcal{B})$. If there exists an automorphism $\sigma$ of order $v$, then the BIBD is called \emph{cyclic} and is denoted by $\mbox{CBIBD}(v,k,\lambda)$. In this case, the point-set $X$ can be identified with $\mathbb{Z}_v$, the set of integers modulo $v$, and $\sigma$ can be represented by $\sigma : i \rightarrow i + 1$ (mod $v$).
For a block $B=\{b_1, \ldots, b_k\}$ in a $\mbox{CBIBD}(v,k,\lambda)$, the set $B + i := \{b_1+i$ (mod $v), \ldots, b_k+i$ (mod $v)\}$ for $i \in \mathbb{Z}_v$ is called a \emph{translate} of $B$, and the set of all distinct translates of $B$ is called the \emph{orbit} containing $B$. If the length of an orbit is $v$, then the orbit is said to be \emph{full}, otherwise \emph{short}. A block chosen arbitrarily from an orbit is called a \emph{base block} (or \emph{starter block}). If $k$ divides $v$, then the orbit containing the block
$B=\big\{0, \frac{v}{k}, 2 \frac{v}{k}, \ldots ,(k-1)\frac{v}{k}\big\}$ is called a \emph{regular short orbit}. For a $\mbox{CBIBD}(v,k,1)$ to exist, a necessary condition is $v \equiv 1$ or $k$ (mod $k(k-1)$). When $v \equiv 1$ (mod $k(k-1)$) all orbits are full, whereas if $v \equiv k$ (mod $k(k-1)$) one orbit is the regular short orbit and the remaining orbits are full.

A BIBD is said to be \emph{resolvable}, and denoted by $\mbox{RBIBD}(v,k,\lambda)$, if the block-set $\mathcal{B}$ can be partitioned into classes $\mathcal{R}_1,\ldots , \mathcal{R}_r$ such that every point of $X$ is contained in exactly one block of each class. The classes $\mathcal{R}_i$ are called \emph{resolution} (or \emph{parallel}) \emph{classes}.
If $\mathcal{R}_i$ is a resolution class, define $\mathcal{R}^\sigma_i:=\{B^\sigma : B \in \mathcal{R}_i\}$. An RBIBD is called \emph{cyclically resolvable} if it has a non-trivial automorphism $\sigma$ of order $v$ that preserves its resolution $\{\mathcal{R}_1,\ldots \mathcal{R}_r\}$, i.e., $\{\mathcal{R}^\sigma_1,\ldots \mathcal{R}^\sigma_r\}= \{\mathcal{R}_1,\ldots \mathcal{R}_r\}$ holds. If, in addition, the design is cyclic with respect to the same automorphism $\sigma$, then it is called  \emph{cyclically resolvable cyclic}, and denoted by $\mbox{CRCBIBD}(v,k,\lambda)$.

In a $\mbox{CBIBD}(v,k,1)$, we can define a multiset $\Delta B :=\{b_i - b_j : i,j = 1, \ldots, k; i \neq j \}$ of \emph{differences} for a base block $B=\{b_1,\ldots,b_k\}$. Let $\{B_i\}_{i \in I}$, for some index set $I$, be all the base blocks of full orbits. If $v \equiv 1$ (mod $k(k-1)$), then clearly $\bigcup_{i \in I} \Delta B_i = \mathbb{Z}_v\setminus\{0\}$. The family of base blocks $\{B_i\}_{i \in I}$ is then called a \emph{(cyclic) difference family} in $\mathbb{Z}_v$, denoted by $\mbox{CDF}(v,k,1)$.

Let $k$ be an odd positive integer, and $p \equiv 1$ (mod $k(k-1)$) a prime. A $\mbox{CDF}(p,k,1)$ is said to be \emph{radical}, and denoted by $\mbox{RDF}(p,k,1)$, if each base block is a coset of the $k$-th roots of unity in $\mathbb{Z}_p$~\cite{Bur95radical}.

\begin{table*}[!t] 
\caption{Possible exceptions: An $\mbox{RBIBD}(v,k,1)$ with $k=5$ or $8$ is not known to exist for the following values of \mbox{$v \equiv k$ (mod $k(k-1)$)}.}\label{table_RBIBD}

\begin{center}
\begin{tabular}{lcccccccc|c|}
 \cline{2-10} 
  & \multicolumn{9}{|c|}{$k=5$} \\ 
\hline
\multicolumn{1}{|c|}{\multirow{1}{*}{$v$}}  & 45 & 345 & 465 & 645 & & & &\\
\hline
 \\
 \cline{2-10} 
  & \multicolumn{9}{|c|}{$k=8$} \\ 
\hline
\multicolumn{1}{|c|}{\multirow{9}{*}{$v$}}  & 176  & 624  & 736  & 1128 & 1240 & 1296 & 1408 & 1464\\
 \multicolumn{1}{|c|}{\multirow{1}{*}{}}  &  1520 &1576  & 1744 & 2136 & 2416 & 2640 & 2920 & 2976\\
  \multicolumn{1}{|c|}{\multirow{2}{*}{}} &  3256 & 3312 & 3424 & 3760 & 3872 & 4264 & 4432 & 5216\\
 \multicolumn{1}{|c|}{\multirow{3}{*}{}} &   5720 & 5776 & 6224 & 6280 & 6448 & 6896 & 6952 & 7008\\
 \multicolumn{1}{|c|}{\multirow{4}{*}{}} &   7456 & 7512 & 7792 & 7848 & 8016 & 9752 & 10200 & 10704\\
 \multicolumn{1}{|c|}{\multirow{6}{*}{}} &   10760 & 10928 & 11040 & 11152 & 11376 & 11656 & 11712 & 11824\\
 \multicolumn{1}{|c|}{\multirow{7}{*}{}} &   11936 & 12216 & 12328 & 12496 & 12552 & 12720 & 12832 & 12888\\
 \multicolumn{1}{|c|}{\multirow{8}{*}{}} &   13000 & 13280 & 13616  & 13840 & 13896 & 14008 & 14176 & 14232\\
  \multicolumn{1}{|c|}{\multirow{9}{*}{}} &  21904 & 24480 &  &  & & & & \\
   \hline
\end{tabular}
\end{center}
\end{table*}

\subsection{BIBD LDPC Code Design}

A (binary) LDPC code~\cite{gal62} is a linear block code with a sparse parity-check matrix (over $GF(2)$). If the parity-check matrix has both constant column and constant row weights, then the code is said to be \emph{regular}, otherwise  \emph{irregular}. We summarize the state-of-the-art of general LDPC code design from BIBDs with respect to our further purposes (cf.~\cite{Kou01,Hon04,Vas04, JohnDiss,Well03, Tan81}):
The $v \times b$ incidence matrix of a $\mbox{BIBD}(v,k,1)$ defines a parity-check matrix $H$ of a binary $(k,r=bk/v)$-regular \emph{BIBD-LDPC code} of length $b=(v(v-1))/(k(k-1))$, if we identify the $v$ points with the parity-check equations and the $b$ blocks with the code bits. In this case, all column weights of $H$ are equal to $k$, and all row weights are equal to $r$. The dimension of the code is $K=b- \text{rank}_2(H)$, where $\text{rank}_2(H)$ denotes the 2\text{-rank} of $H$, and the code rate is $R=(b - \text{rank}_2(H))/b$. 
In general, often the calculation of $\text{rank}_2(H)$ depends on the specific structure of the $\mbox{BIBD}(v,k,1)$.
It is clear from the classical Fisher's Inequality that $\text{rank}_2(H) \leq v$. Furthermore, it is elementary to see that, if $2 \mid \frac{v-1}{k-1}$, then $\text{rank}_2(H) \geq v-1$, with equality if and only if $2 \mid k$ (cf.~\cite[Thm.\,2.4.1]{Assm92}). More precise results are in particular known for $\mbox{BIBD}(v,3,1)$s~\cite{Doy78}.

As in a $\mbox{BIBD}(v,k,1)$ no pair of points have more than one block in common (i.e., no pair of columns of $H$ contains more than one ``1'' at the same positions), the Tanner graph of a BIBD-LDPC code is free of cycles of length 4 and hence has girth at least 6. 
Moreover, the fact that every pair of points occurs in exactly one block ensures that the girth is exactly 6. Since each column in $H$ has weight $k$ and no two columns have more than one ``1'' in common, there are $k$ parity-check equations that are orthogonal on every code bit (i.e., the code is one-step majority logic decodable). Thus, the minimum distance of the code is at least $k+1$. 
The same lower bound holds for the size of minimal stopping sets~\cite{Kash03}.

Some specific classes of BIBDs proved to be especially useful in the code design:

\begin{enumerate}

\item[(a)] LDPC code constructions based on CDFs have been considered in~\cite{Hon04,Vas04,JohnDiss} for constant column weights $k= 3$ to $7$. CDF-LDPC codes have the structural advantage that each orbit of a base block in the BIBD corresponds to a circulant submatrix in the parity-check matrix. Hence, these codes a have quasi-cyclic structure (cf.~\cite{Town}), which allows linear-complexity encoding with simple feedback shift registers.

\item[(b)] LDPC codes based on RBIBDs have been constructed in~\cite{JohnDiss,Well03} for constant column weights $k= 3,4$. RBIBD-LDPC codes have the property that they have very sparse parity-check matrices.

\item[(c)] LDPC codes based on CRCBIBDs have been designed in~\cite{JohnDiss,Well03} for constant column weights  $k= 3,4$. CRCBIBD-LDPC codes 
exhibit very sparse parity-check matrices and, in addition, their structure allows linear-complexity encoding with simple feedback shift registers similarly as for quasi-cyclic codes.

\end{enumerate}

\begin{table*}[!t] 
\caption{Existence of a $\mbox{CRCBIBD}(pk,k,1)$ with $k=5$, $p<10^3$, and $k=7$ or $9$, $p < 10^4$.}\label{table_RDF}
\begin{center}
\begin{tabular}{lccccccccc|}
 \cline{2-10} 
  & \multicolumn{9}{|c|}{$k=5$} \\ 
\hline
\multicolumn{1}{|c|}{\multirow{2}{*}{$p$}}  & 41 & 61 & 241 & 281 & 401 & 421 & 601 & 641 & 661 \\ 
\multicolumn{1}{|c|}{\multirow{3}{*}{}}   & 701 & 761 & 821 & 881 &  &  &  &    &  \\    
\hline
\\   
\cline{2-10}
  & \multicolumn{9}{|c|}{$k=7$} \\   
  \hline
\multicolumn{1}{|c|}{\multirow{3}{*}{$p$}}  &   337  & 421  &  463 & 883 & 1723 & 3067 & 3319 & 3823 & 3907 \\
\multicolumn{1}{|c|}{\multirow{3}{*}{}}   &   4621 & 4957 & 5167 & 5419 & 5881 & 6133 & 8233 & 8527 & 8821 \\
\multicolumn{1}{|c|}{\multirow{3}{*}{}}   &   9619 & 9787 & 9829 & & & & & & \\
   \hline
\\
\cline{2-10}
  & \multicolumn{9}{|c|}{$k=9$} \\   
  \hline
\multicolumn{1}{|c|}{\multirow{1}{*}{$p$}}  & 73  & 1153  & 1873  & 2017 & 6481 & 7489 & 7561 &  &  \\
   \hline
\end{tabular}
\end{center}
\end{table*}

\section{New LDPC Codes with Higher Column Weights}\label{new}

In this section, we extend the previous results on LDPC codes based on CDFs, RBIBDs and CRCBIBDs to higher column weights.
Experimental results on decoding performance with the sum-product algorithm show that the novel  LDPC codes offer substantial practical application potential (Subsection~\ref{sim}).

\subsection{CDF LDPC Codes}

\begin{theorem}\label{thm_CDF-LDPC Codes}
Let $p$ be a prime. Then there exists a $(k,r)$-regular quasi-cyclic LDPC code of length $pr/k$ and rate at least $(r-k)/r$ based on a $\mbox{CDF}(p,k,1)$ for the following cases:

\begin{enumerate}

\item[(1)] $(k,p)= (3,6t+1)$ for any positive integer $t$,

\item[(2)] $(k,p)= (4,12t+1)$ for any positive integer $t$,

\item[(3)] $(k,p)= (5,20t+1)$ for any positive integer $t$,

\item[(4)] $(k,p)= (6,30t+1)$ for any positive integer $t\neq 2$,

\item[(5)] $(k,p)= (7,42t+1)$ for any positive integer $t>1$ with the possible exceptions $p=127, 211$ as well as primes $p \in [261239791, 1{.}236597 \times 10^{13}]$ such that $(-3)^{\frac{q-1}{14}} = 1$ in $\mathbb{Z}_p$.

\item[(6)] $(k,p)= (8,p)$ for all values of $p \equiv 1$ (mod $56$) $ < 10^4$ with the possible exceptions $p=113, 169, 281, 337$,

\item[(7)] $(k,p)= (9,p)$ for all values of $p \equiv 1$ (mod $72$) $ < 10^4$ with the possible exceptions $p=289, 361$.

\end{enumerate}
\end{theorem}

\begin{IEEEproof}
The proof relies on known infinite ((1)-(5)) and finite ((6)-(7)) families of  $\mbox{CDF}(p,k,1)$'s (cf.~\cite{CWZ02} and the references therein;~\cite{crc06}).
\end{IEEEproof}

We note that Cases $(1)$--$(5)$ have been presented in~\cite{Hon04,Vas04,JohnDiss}.

\begin{example}
A $(5,26)$-regular quasi-cyclic LDPC code of length 546 and rate at least 0.81 can be obtained from a $\mbox{CDF}(105,5,1)$.
\end{example}

\subsection{RBIBD LDPC Codes} 

\begin{theorem}\label{thm_RBIBD-LDPC Codes}
Let $v$ be a positive integer. Then there exists a $(k,r)$-regular LDPC code of length $vr/k$ 
and rate at least $(r-k)/r$ based on an $\mbox{RBIBD}(v,k,1)$ for the following cases:

\begin{enumerate}

\item[(1)] $(k,v)= (3,6t+3)$ for any positive integer $t$,

\item[(2)] $(k,v)= (4,12t+4)$ for any positive integer $t$,

\item[(3)] $(k,v)= (5,20t+5)$ for any positive integer $t$ with the possible exceptions given in Table~\ref{table_RBIBD},

\item[(4)] $(k,v)= (8,56t+8)$ for any positive integer $t$ with the possible exceptions given in Table~\ref{table_RBIBD}.

\end{enumerate}

\end{theorem}

\begin{IEEEproof}
The proof is based on known infinite series of $\mbox{RBIBD}(v,k,1)$'s (cf.~\cite{GA97,crc06} and the references therein).
\end{IEEEproof}

We remark that Cases $(1)$ and $(2)$ of have been considered in~\cite{JohnDiss,Well03}.

\begin{example}
A $(5,46)$-regular LDPC code of length 1702 and rate at least 0.89 can be constructed from an $\mbox{RBIBD}(185,5,1)$.
\end{example}

\begin{theorem}\label{thm_RBIBD-LDPC Codes_prime}
If $v$ and $k$ are both powers of the same prime, then a $(k,r)$-regular LDPC code of length $vr/k$ and rate at least $(r-k)/r$ based on an $\mbox{RBIBD}(v,k,1)$ exists if and only if $(v-1) \equiv 0$ (mod $(k-1)$) and $v \equiv 0$ (mod $k$).
\end{theorem}

\begin{IEEEproof}
It has been shown in~\cite{Greig06} that, for $v$ and $k$  both powers of the same prime, the necessary conditions for the existence of an $\mbox{RBIBD}(v,k,\lambda)$ are sufficient.
\end{IEEEproof}

\subsection{CRCBIBD LDPC Codes}

\begin{theorem}\label{thm_CRCBIBD-LDPC Codes}
Let $p$ be a prime. Then there exists a $(k,r)$-regular LDPC code of length $pr$ and rate at least $(r-k)/r$ based on a $\mbox{CRCBIBD}(pk,k,1)$ for the following cases:

\begin{enumerate}

\item[(1)] $(k,p)= (3,6t+1)$ for any positive integer $t$,

\item[(2)] $(k,p)= (4,12t+1)$ for any odd positive integer $t$,

\item[(3a)] $(k,p)= (5,20t+1)$ for any positive integer $t$ such that $p<10^3$, and furthermore

\item[(3b)] $(k,p)= (5,20t+1)$ for any positive integer $t$ satisfying the following condition:  $\varepsilon +1$ is not a $2^{e+1}$-th power in $\mathbb{Z}_p$, or equivalently $(11+5\sqrt{5})/2$ is not a $2^{e+1}$-th power in $\mathbb{Z}_p$, where $2^e$ is the largest power of $2$ dividing $t$ and $\varepsilon$ a $5$-th primitive root of unity in $\mathbb{Z}_p$,

\item[(4)] $(k,p)= (7,42t+1)$ for any positive integer $t$ satisfying the following condition:  there exists an integer $f$ such that $3^f$ divides $t$ and $\varepsilon +1, \varepsilon^2 + \varepsilon +1,\frac{\varepsilon^2 + \varepsilon +1}{\varepsilon +1}$ are $3^{f}$-th powers but not $3^{f+1}$-th powers in $\mathbb{Z}_p$, where $\varepsilon$ is a $7$-th primitive root of unity in $\mathbb{Z}_p$,

\item[(5)] $(k,p)= (9,p)$ for the values of $p \equiv 1$ (mod $72$) $ < 10^4$ given in Table~\ref{table_RDF}.

\end{enumerate}

Moreover, there exists a $(k,r)$-regular LDPC code of length $\ell r$ and rate at least $(r-k)/r$ based on a $\mbox{CRCBIBD}(\ell k,k,1)$
for the following cases:

\begin{enumerate}

\item[(6)] $(k,\ell)$ for $k=3,5,7$, or $9$, and $\ell$ is a product of primes of the form $p \equiv 1$ (mod $k(k-1)$) as in the cases above,

\item[(7)] $(k,\ell)= (4,\ell)$ and $\ell$ is a product of primes of the form $p=12t+1$ with $t$ odd.

\end{enumerate}

\end{theorem}

\begin{IEEEproof}
The proof is given in Appendix~\ref{APP_thm_CRCBIBD-LDPC Codes}.
\end{IEEEproof}

We note that Cases $(1)$ and $(2)$ have been  treated in~\cite{JohnDiss,Well03}.

\begin{example}
A $(5,51)$-regular LDPC code of length 2091 and rate at least 0.90 can be obtained from a $\mbox{CRCBIBD}(41\times5,5,1)$.  
\end{example}

\subsection{Simulation Results}\label{sim}

\begin{figure*}[!ht]
	\centerline{
		\subfloat{
			\includegraphics[scale=0.7, trim = 0cm 0cm 0cm 1cm]{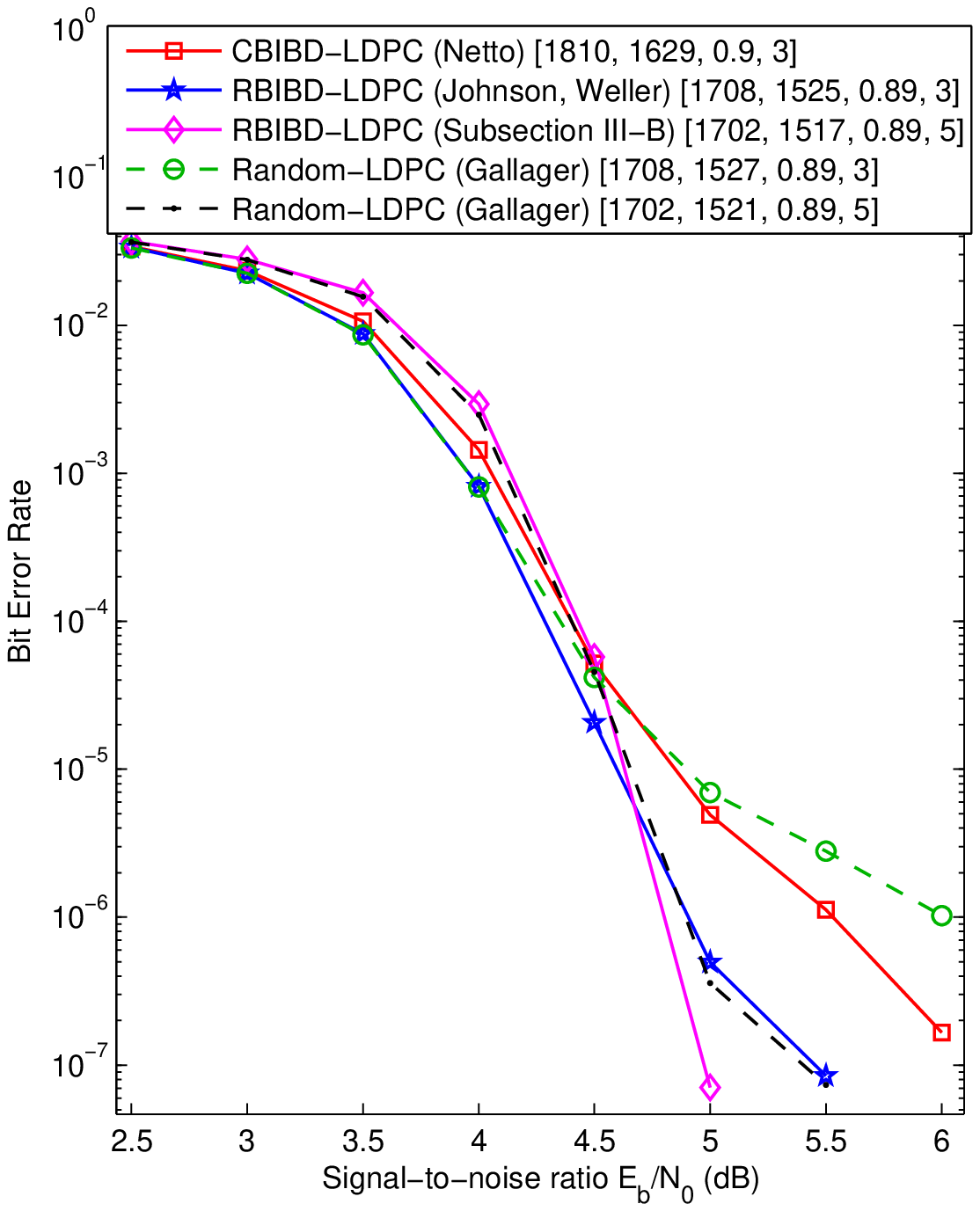}
		}
		\hfill
		\subfloat{
			\includegraphics[scale=0.7, trim = 0cm 0cm 0cm 1cm]{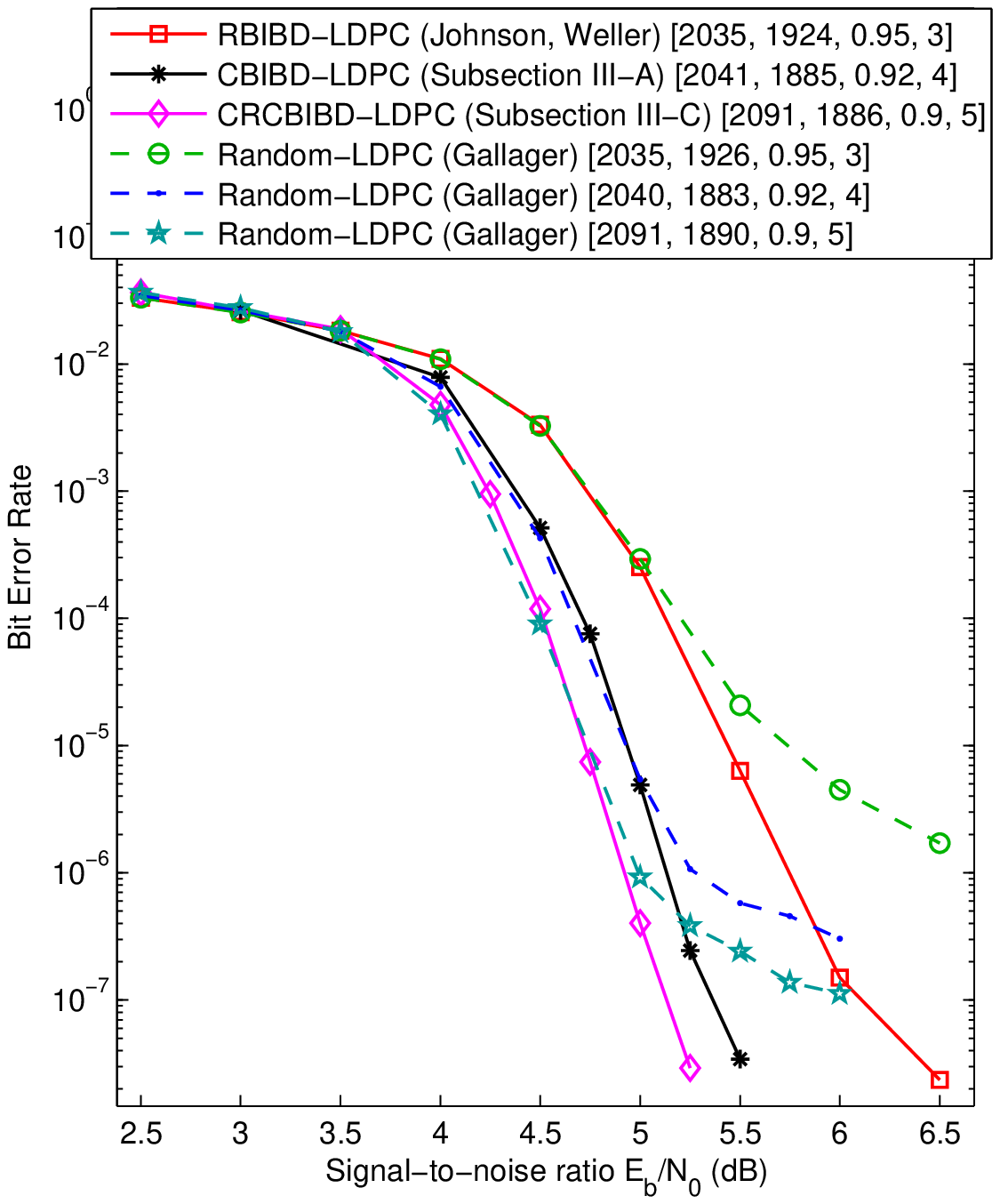}
		}
	}
	\caption{Decoding performance of various novel structured LDPC codes}
	\label{table_Simu1}
\end{figure*}

For experimental results on decoding performance, we employed the iterative sum-product algorithm, as proposed in~\cite{mncN}, on an AWGN channel with a maximum of 50 iterations per codeword. Fig.~\ref{table_Simu1} shows the bit-error rate (BER) performance of LDPC codes that have been constructed using the different combinatorial techniques presented in this section, and random regular Gallager LDPC codes~\cite{gal62}. A legend displays the following information in the respective order: code type, construction method in brackets, and a quadruple $[N, K, R, k]$ consisting of block length $N$, dimension $K$, code rate $R$ and column weight $k$ of the parity-check matrix.

The left plot of Fig.~\ref{table_Simu1} shows, that codes with higher column weights may have a relatively small performance deficit at low signal-to-noise-ratios (SNRs), but outperform their counterparts for higher SNRs significantly. The CBIBD-LDPC code based on Netto's construction (cf.~\cite{Hon04,Vas04,JohnDiss}) and the RBIBD-LDPC code of Johnson and Weller (cf.~\cite{JohnDiss,Well03}) have been truncated in length to yield parameters comparable to those of our new RBIBD-LDPC code.
Both plots demonstrate that the new structured codes outperform known random Gallager LDPC codes and known structured BIBD-LDPC codes with comparable parameter sets. We observe a performance gain as the column weights grow larger and a particular good performance of CRCBIBD-LDPC codes.

\section{New weight-$q$ RA Codes}\label{new2}

As an extension of the accumulator type for w3RA codes introduced by Johnson and Weller~\cite{combinatorial_interleavers_for_systematic_ra_codes}, we present a generalized accumulator for wqRA codes. This accumulator structure was similarly introduced by Liva et al.~\cite{Liva05} with focus on the design of irregular RA codes. The intention of their work is to allow a more flexible choice of the nodes degree distribution for optimizing the code in the sense of density evolution~\cite{Urb08}, while we use the accumulator to provide new sRA codes with an improved decoding performance closer to those of regular LDPC codes. Let $g_1,\ldots,g_{q-1}\in \{1,...,M\}$ denote the design parameters of the generalized accumulator, where $g_i\neq g_j$ for $i\neq j$,  and $s_l := \sum_{j=1}^l g_j$, $l\in \{1,\ldots,q-1\}$. Let $r_i$ be the output of the combiner at time $i$. The output of the accumulator $p_i$ is now
\[
\begin{array}{lcll}
p_i & := & r_i, &i=1,...,s_1\\
p_i & := & r_i \oplus p_{i-s_1}, &i=s_1+1,...,s_2\\
    & \vdots & &\ \ \; \vdots\\
p_i & := & r_i \oplus p_{i-s_1} \oplus ... \oplus p_{i-s_{q-1}},&i=s_{q-1}+1,...,M.\\
\end{array}
\]

This scheme again results in a parity-check matrix of the form $H = [H_1 H_2]$, where the $i$-th column of $H_2$ has \mbox{1-entries} at rows $i$ and $i+s_l$, $1\leq l \leq q-1$, given that these rows exist.  Thus, the design parameters $g_j$ specify the vertical distance between the \mbox{$j$-th} and \mbox{$(j+1)$-th} \mbox{1-entry} in the columns of $H_2$. The $H_1$-part is unaffected by the accumulator and hence remains as introduced in Section \ref{intro}. Consequently, most of the columns of $H_2$ have the same (high) column weight $q$ as the columns of $H_1$, leading to an improved decoding performance compared to sRA codes.

\subsection{CBIBD sRA Codes}

\begin{figure*}[!t] \centerline{
		\subfloat[CBIBD$(19,3,1)$ sRA code]{
			\includegraphics[width=3in]{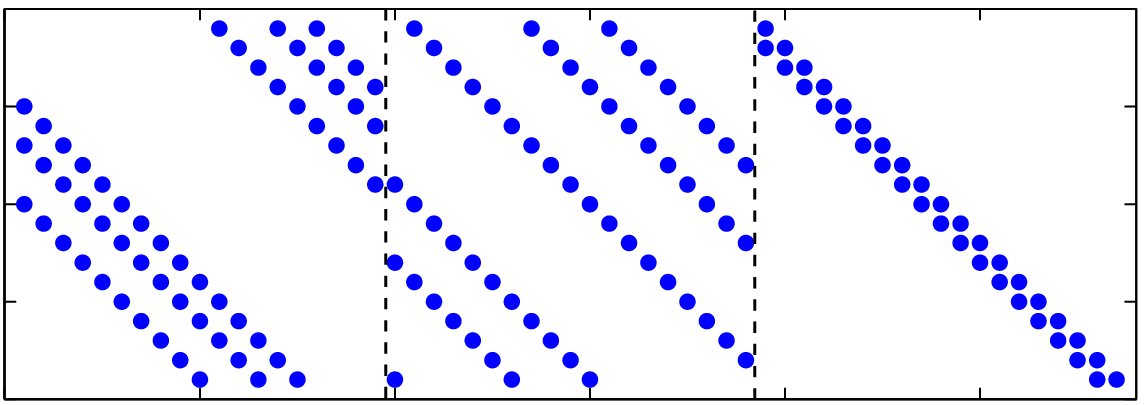}
			\label{cbibd}
		}
		\hfil
		\subfloat[RBIBD$(21,3,1)$ sRA code]{
			\includegraphics[width=3in]{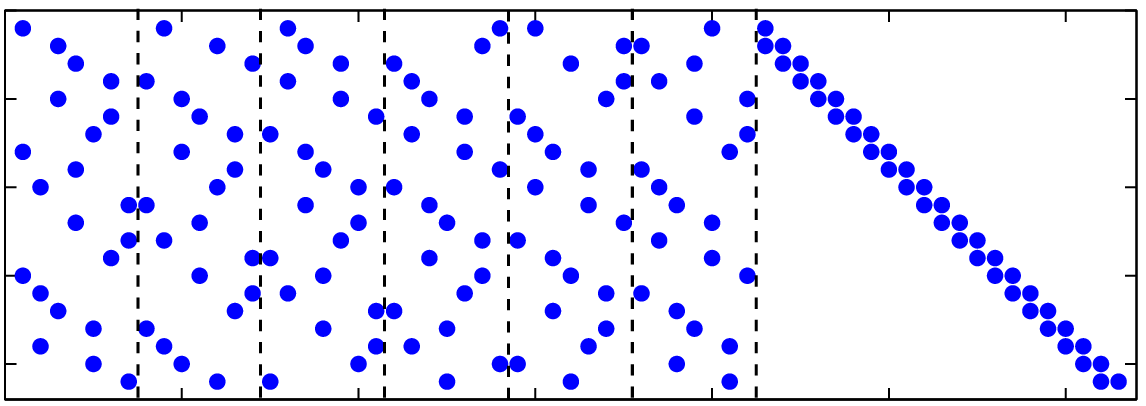}
			\label{kts_parity_check}
		}
	}
	\caption{Parity-check matrices of two systematic RA codes}
\end{figure*}

The incidence matrix of a CBIBD has a quasi-cyclic structure $A=[A_1 A_2 , ..., A_l]$ of circulant submatrices $A_i$ and can easily be transformed into the parity-check matrix of an sRA code. For this purpose, we consider the underlying difference family. A CDF $F$ always contains a base block $B_t \in F$ with $1\in \Delta B_t$. The translates of $B_t$ give rise to the circulant submatrix $A_t$, where every column has two consecutive \mbox{1-entries}. We retain all pairs of ones and delete the others to obtain the required double diagonal matrix $H_2$ for sRA codes. Finally, we can use an arbitrary number of the remaining $A_i$, $1\leq i\leq l,\ i\neq t$ to form $H_1$ (Fig. \ref{cbibd}).

It is also possible to construct weight-$q$ RA codes from CBIBDs, where one of the design parameters $g_1,...,g_{q-1}$ can be chosen arbitrarily, say $g_1$ w.l.o.g. We have to identify the base block $B_s$ with $g_1 \in \Delta B_s$ in the underlying CDF and transform the associated $A_s$ into the accumulator matrix $H_2$ by simply deleting the 1-entries of the upper triangular. The other design parameters arise from the remaining differences of $B_s$. 

\subsubsection{Netto's Construction for column weights 3}

Let $GF(p)$ be a Galois field of prime order $p=6t+1$ and $G$ its multiplicative group. Let $\omega$ be a primitive element of the field. Then the set \mbox{$F:=\{\omega^i G^{2t} \bmod p \ |\ 1\leq i \leq t\}$} is a CDF$(6t+1,3,1)$~\cite{netto}.

\subsubsection{New sRA and w3RA Codes}\label{subB}

With Netto's construction we can obtain both sRA codes and w3RA codes. For this purpose, we have to find the base block $B_s = \{\omega^s, \omega^{s+2t}, \omega^{s+4t}\}$ (mod $p$), $s \in \{1,...,t\}$, that generates the differences 1 and $g_1$, respectively. The position of $B_s$ depends on the distribution of the primitive element $\omega$ and it can be shown that this search problem is equivalent to solving the discrete logarithm (see Appendix~\ref{DiscreteLog}). Thus, finding the required base block in general is difficult for large difference families. We may now proceed to construct the desired sRA codes and w3RA codes as described earlier in this section. 

\subsubsection{Buratti's Construction for column weights 4 and 5}

Let $GF(p)$ be a Galois field of prime order $p=12t+1$ and $H$ its multiplicative group. If we find some $b\in GF(p)$, $B=\{0,1,b,b^2\}$ such that $\Delta^+ B:=\{b_i-b_j|1\leq j < i \leq 4\}$ is a representative system of the cosets of $H^6$ in $H$, then $F := \{\omega^{6i}B \bmod p \ |\ 1\leq i \leq t\}$ is a CDF$(p,4,1)$~\cite{Buratti1995}.
Similarly, for $p=20t+1$ we can construct a CDF$(p,5,1)$ by $F := \{\omega^{10i}B \bmod p \ |\ 1\leq i \leq t\}$, if we find some $b\in GF(p)$, $B=\{0,1,b,b^2, b^3\}$ such that $\Delta^+ B$ is a representative system of the cosets of $H^{10}$ in $H$~\cite{Buratti1995}.

\subsubsection{New sRA, w4RA and w5RA Codes}

For obtaining sRA codes or wqRA codes with $g_1=1$, we have to identify the base block $B_i \in F$ with $1 \in \Delta B_i$. For Buratti's construction we can show that the desired block is the last one of the family given by $i=t$.
\begin{IEEEproof}
It holds that $\omega^{(p-1)/2} \bmod p = p-1$. In case of $p=12t+1$, the last block contains $0$ and $\omega^{6t} \bmod p = p-1$. These two elements generate the required difference $0-(p-1) \bmod p = 1$. In case of $p=20t+1$, the proof is analogous. 
\end{IEEEproof}

Therefore, we can efficiently derive sRA codes or wqRA codes with $g_1=1$ from the underlying difference family in contrast to the construction in Subsection~\ref{subB}. 

\subsection{RBIBD sRA Codes}

Let $B(i)$ be the column of a matrix which contains consecutive ones at position $i$ and $i+1$. For sRA codes, we have to identify all $B(i)$, $1\leq i \leq M-1$, in the incidence matrix of an RBIBD for constructing the double diagonal matrix $H_2$. Unfortunately, these columns are spread over different resolution classes and thus, taking these columns for $H_2$ would destroy several classes. Therefore, we propose to build up $H_2$ only from a few resolution classes and let the others unaffected (Fig. \ref{kts_parity_check}). The advantage of this approach is that we can use an arbitrary number of the intact resolution classes to form $H_1$ and hence have more flexibility in the code design. It is important to omit only complete resolution classes, because we then remove a definite number of ones per row which ensures the regularity of the resulting parity-check matrix.

\subsubsection{sRA Codes by Johnson and Weller}

Johnson and Weller rely on the construction of Ray-Chaudhuri and Wilson~\cite{chaudhuri_wilson} for generating an $\mbox{RBIBD}(v,3,1)$, also called a Kirkman triple system (KTS). The last three resolution classes of the KTS have a special structure consisting of circulant submatrices. These classes can be transformed into a double diagonal matrix for sRA codes with simple row and column permutations~\cite{combinatorial_interleavers_for_systematic_ra_codes}. 
More precisely, let \[
C = \left[
\begin{array}{ccc}
I^a & I^d & I^g\\
I^b & I^e & I^h\\
I^c & I^f & I^i\\
\end{array}
\right]
\]
be the last three resolution classes, where every class is vertically divided into three circulant submatrices $I^i$, which are cyclic shifts by $i$ places of the $M/3\times M/3$ identity matrix. Johnson and Weller first replace $I^c, I^d$ and $I^h$ with all zeros matrices and then permute it into a double diagonal matrix, where the row permutations must be applied to the entire incidence matrix. Finally, an arbitrary number of the remaining resolution classes can be used to form $H_1$.

\subsubsection{Modification for New w3RA Codes}

We propose here not to replace the submatrices  $I^c, I^d$ and $I^h$ with zero matrices, but to keep the entries in $C$. With the same permutations we obtain the parity-check matrix of a w3RA code, where $g_1=1$ and $g_2$ arise from the construction.

\subsection{CRCBIBD sRA Codes}

Codes from CRCBIBDs combine the resolvability of RBIBDs and the cyclic character of CBIBDs, so that we gain the specific advantages of both code classes. The cyclic structure enables linear-complexity encoding, while we have the flexibility of RBIBDs to adjust the length and rate of the code independently by removing an arbitrary number of resolution classes. Although the resulting CRCBIBD codes are not quasi-cyclic, we may realize linear-complexity encoding either by a modified scheme of feedback shift registers as proposed in~\cite{JohnDiss} or by employing the encoder of an sRA code. For the latter, there is no need to maintain the quasi-cyclic structure of the parity-check matrix and we thus may use an arbitrary selection of the resolution classes for designing the code.

\subsubsection{Construction of Genma, Mishima and Jimbo}

With the construction of Genma, Mishima and Jimbo~\cite{Gen97}, we can obtain a CRCBIBD$(pk,k,1)$ with $p$ prime and $k$ odd, if there exists an RDF$(p,k,1)$. In their proof, they construct all base blocks of the design and use them to partition the design into resolution classes. 
For our purpose to construct sRA codes, we need one of the resolution classes given by $\mathcal{R}'_{j,0}$ and its class orbit
\[
\mathcal{R}'_{j,l}=\mathcal{R}'_{j,0}+l \pmod{pk},\ l\in {\mathbb{Z}}_k,\ j\in I,
\]
where $I$ is a suitable subset of $\{0,1,...,2mn-1\}$.

\begin{figure*}[!t]
	\centerline{
		\subfloat{
			\includegraphics[width=5in]{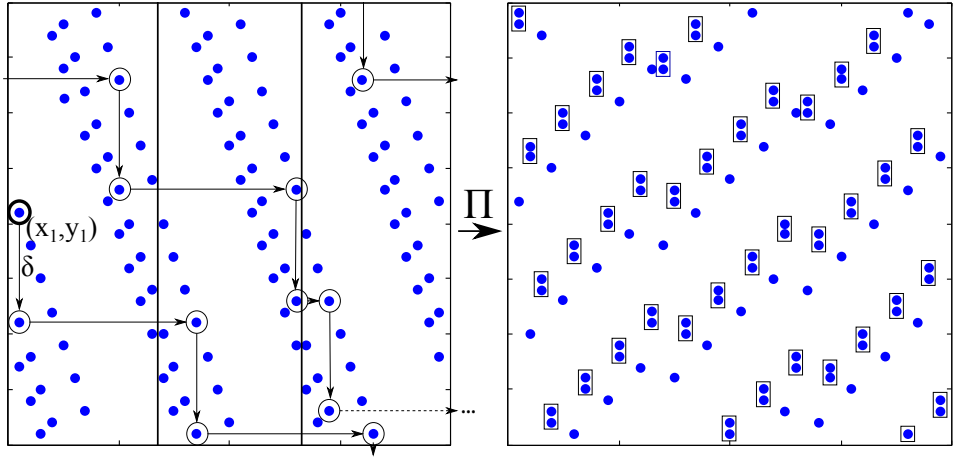}
		}
	}
	\caption{Transformation of $C$ into $\widehat{H_2}$ for a CRCBIBD$(39,3,1)$ code}
	\label{crcbibd_trans}
\end{figure*}

\subsubsection{New sRA and wqRA Codes}

The class orbit $\mathcal{R}'_{j,l}, l\in {\mathbb{Z}}_k$ consists of $k$ resolution classes, which are cyclic shifts of each other. Let $C$ be the incidence matrix of this class orbit. 
Now let $(x_1,y_1)$ be the position of an arbitrary 1-entry in $C$ and $\delta$ be the vertical distance to the next 1-entry in column $y_1$ (Fig. \ref{crcbibd_trans}). It must hold that $\gcd(\delta, pk)=1$, otherwise we have to choose another 1-entry with different $\delta$. Furthermore let $\Pi$ be a permutation vector for the rows of $C$, where $\Pi[i]$ is the value at position $i,\ 1\leq i \leq pk$.
Now we compute
\[
\Pi[i]=x_1+\delta (i-1) \pmod{pk}
\]
and apply the permutation to the rows of the entire incidence matrix. Here we consider only the partial matrix $\widehat{H_2}$ corresponding to $C$ (Fig.~\ref{crcbibd_trans}). For $\widehat{H_2}$ it holds that
\begin{enumerate}
	\item there exists exactly one column $B(i)$ with 1-entries at row $i$ and $i+1$ (mod $pk$) for every $i,\ 1\leq i \leq pk$, and
	\item all $B(i)$ are distinct.
\end{enumerate}

\begin{IEEEproof}
The column $y_1$ of $C$ with 1-entries at rows $x_1$ and $x_1+\delta$ gives the column $B(1)$ of $\widehat{H_2}$. Because of the cyclic resolvability there must be a resolution class with a column $y_{t+1}$ that contains 1-entries at rows $x_1+t\delta$ (mod $pk$) and $x_1+\delta+t\delta$ (mod $pk$), $t\in{1,...,pk-1}$. With $\gcd(\delta,pk)=1$, we have $\delta \mathbb{Z}_{pk} = \mathbb{Z}_{pk}$ and hence we never use a row twice. The columns $y_{t+1}$ result in $B(t+1)$ of $\widehat{H_2}$ and thus we have $B(i)'s$ for every $i$. Moreover, $B(i)$ must be unique, because there can only exist one column with 1-entries at rows $x_1+r\delta$ (mod $pk$) and $x_1+(r+1)\delta$ (mod $pk$), $r\in \{0,...,pk-1\}$, due to the axioms of the design.

The columns $y_i$ must be distinct because the blocks of $\mathcal{R}'_{j,0}$ can never have two pairs of entries with difference $\delta$ due to constructional reasons. Thus, the resulting $B(i)$ must be distinct.
\end{IEEEproof}

Finally, we can transform $\widehat{H_2}$ into a double diagonal matrix by simple column reordering and deleting the superfluous ones. We can even obtain wqRA codes with an arbitrary design parameter $g_1$ with the restriction that $\gcd(g_1,pk)=1$. For this, we modify our computation to
\[
\Pi[i\cdot g_1]=x_1+\delta (i-1) \pmod{pk},\ 1\leq i \leq pk.
\]

\subsection{Simulation Results}\label{sim2}

\begin{figure*}[!ht]
		\centerline{
		\subfloat{
			\includegraphics[scale = 0.65, trim = 0cm 0cm 0cm 1cm]{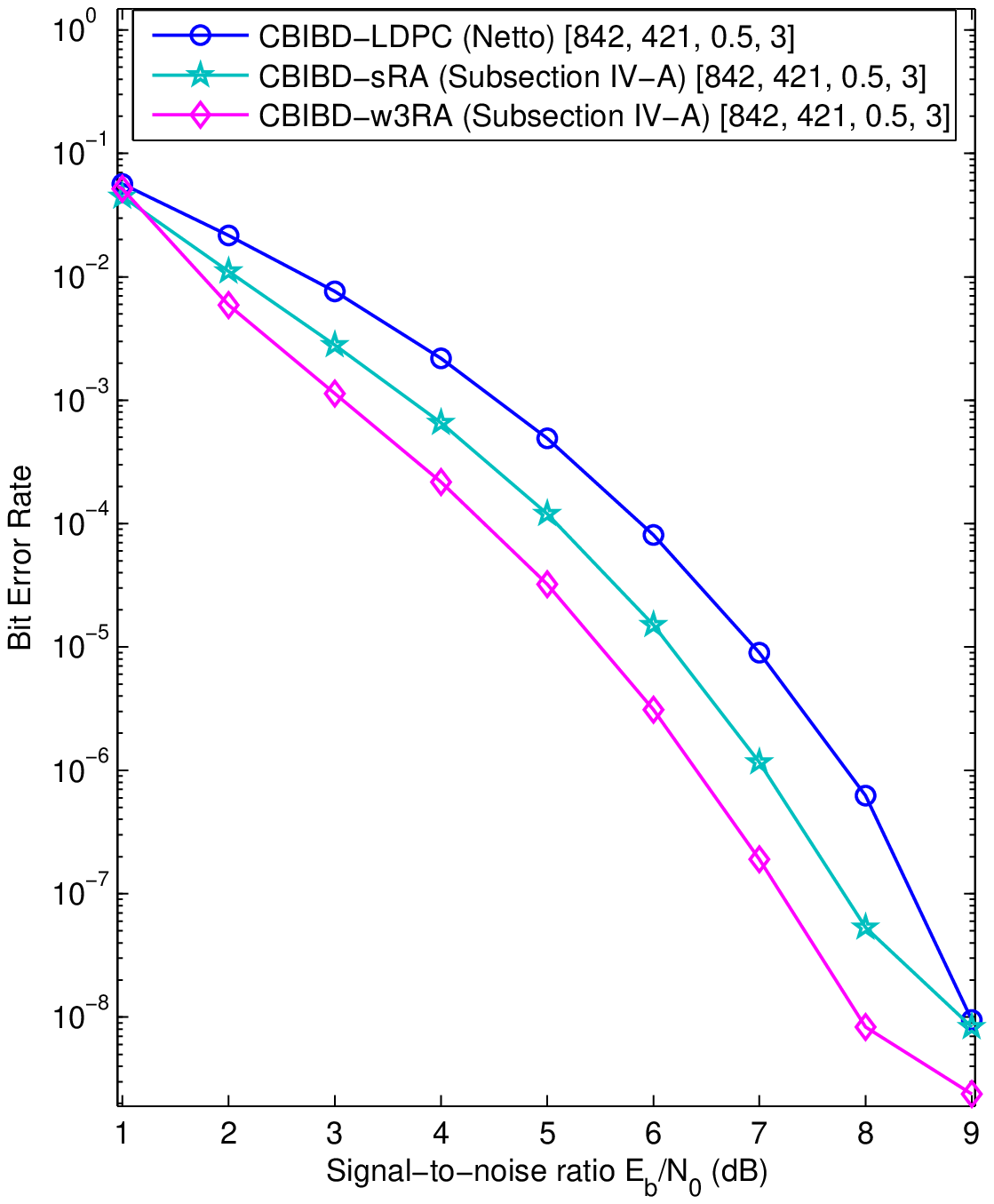}
		}
		\subfloat{
			\includegraphics[scale = 0.65, trim = 0cm 0cm 0cm 1cm]{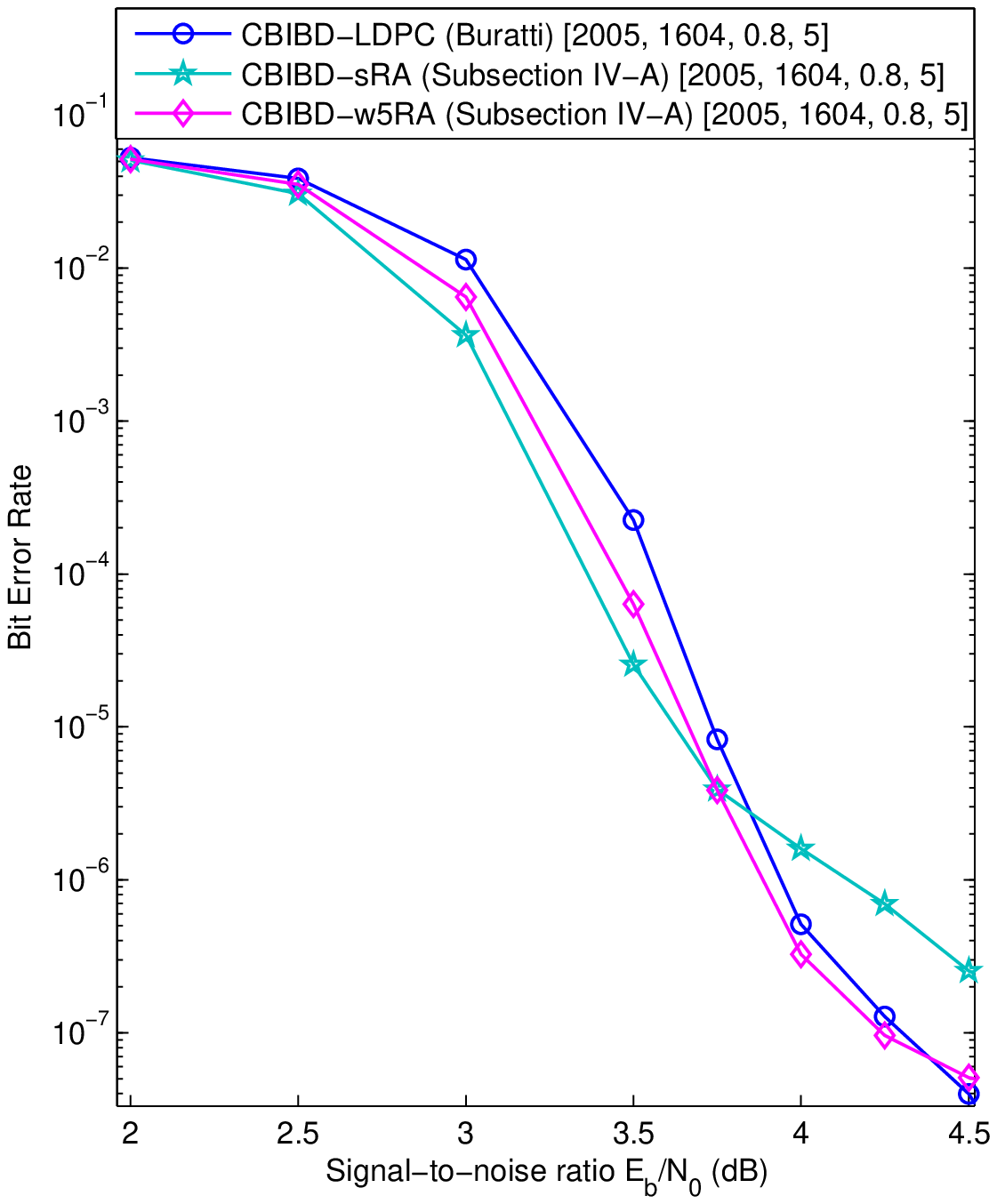}
		}
		}
		
		\centerline{
		\subfloat{
			\includegraphics[scale = 0.65, trim = 0cm 0cm 0cm 1cm]{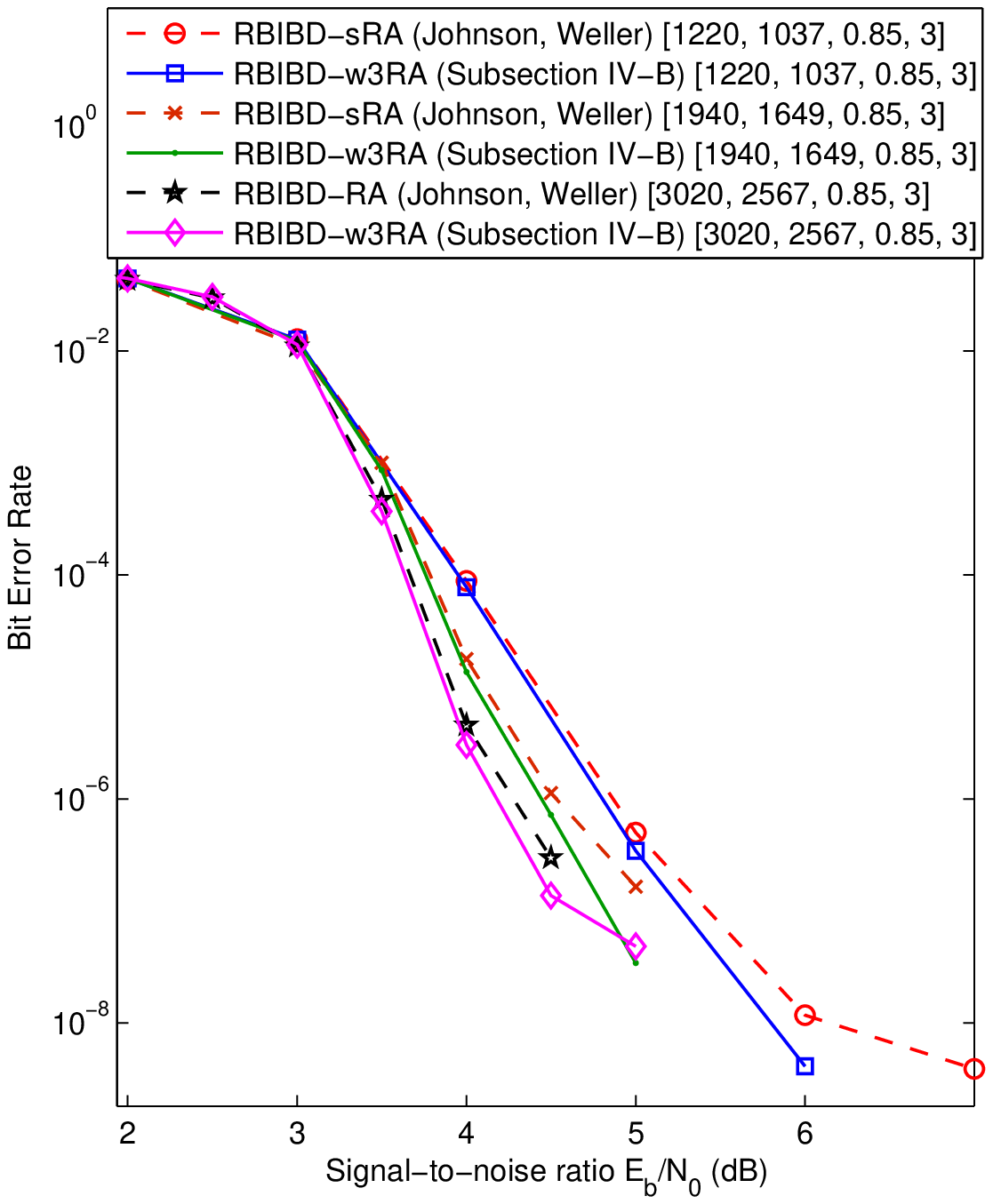}
		}
		\subfloat{
			\includegraphics[scale = 0.65, trim = 0cm 0cm 0cm 1cm]{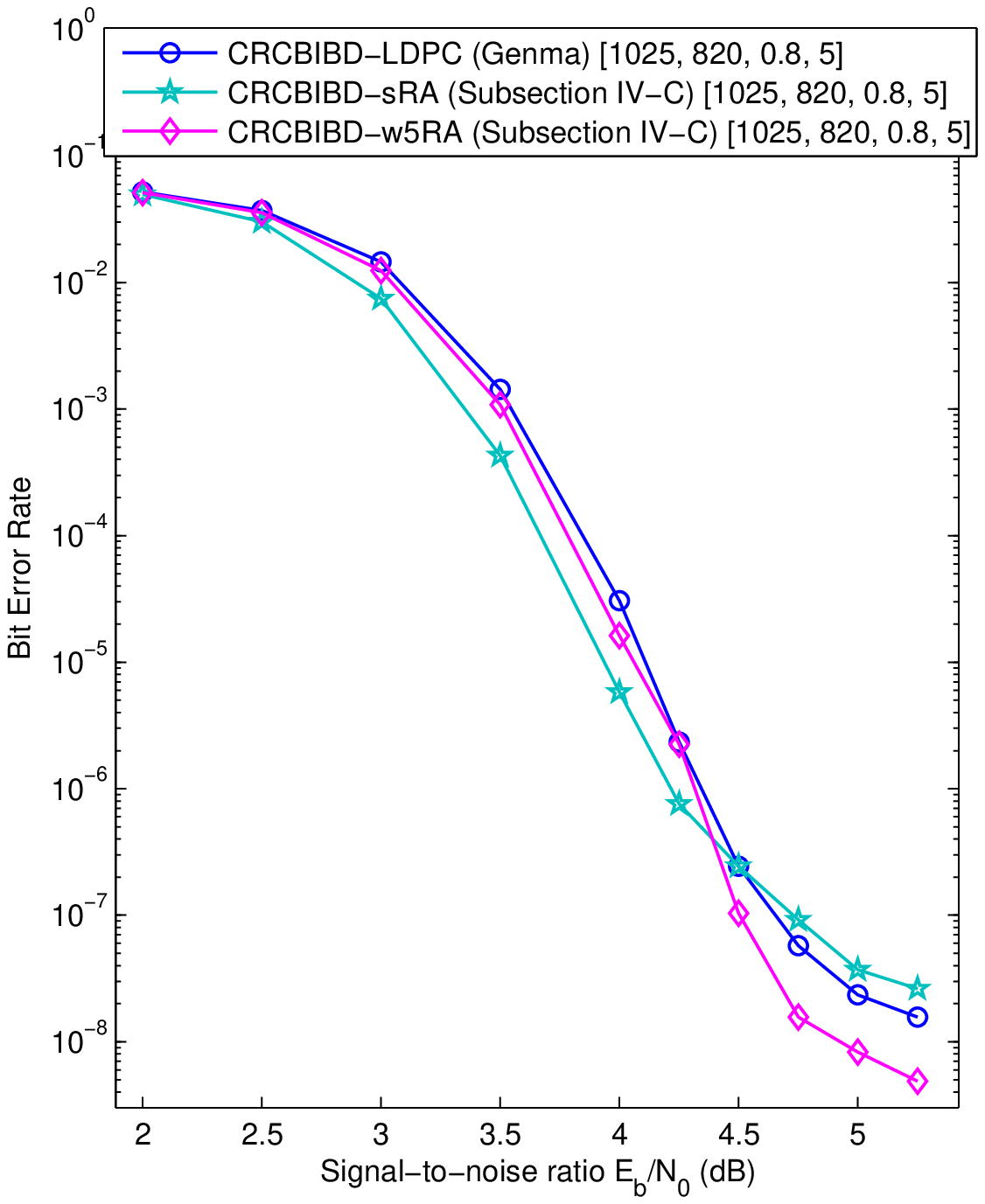}
		}
		}
		\caption{Decoding performance of various novel systematic RA and wqRA codes}
	\label{table_Simu2}
\end{figure*}

In this section, we compare the decoding performance of sRA codes, wqRA codes and regular LDPC codes for our new constructions (Fig.~\ref{table_Simu2}). 
These codes differ solely in the $H_2$-part of their parity-check matrices and thus, the largest differences in performance are expected for short to moderate block lengths or for lower code rates, as in these cases the impact of $H_2$ is relatively higher than that of $H_1$.
For the decoding we use the sum-product algorithm~\cite{mncN} on an AWGN channel with a maximum of 50 iterations per codeword. The LDPC codes are again described by a quadruple $[N, K, R, k]$, and the sRA/wqRA codes by a slightly different quadruple $[N, K, R, q]$, where $q$ is the column weight of $H_1$.
The first plot of Fig.~\ref{table_Simu2} demonstrates the performance gain of a new low-rate w3RA code at relatively small block length, compared to the sRA and LDPC code of the same parameters. In the second plot, our w5RA-code shows a similarly well low-SNR performance as the corresponding sRA code and the excellent high-SNR performance as the LDPC code. The sRA code suffers a high error-floor at approximately 3.75~dB, arising from many weight-2 columns in the parity-check matrix. This error-floor turns out stronger compared to the first plot, since we reduced the column weights of the $H_2$-part from 5 (instead of 3) to 2 in order to obtain a double diagonal matrix. Consequently, the larger decrease of the column weights negatively affects the decoding performance in the error-floor region. The third plot demonstrates the performance gain of our novel w3RA codes compared to the sRA codes of Johnson and Weller, for various codes of rate 0.85 and block lengths varying from 1220 to 3020. The fourth plot shows a particular good high-SNR performance of our w5RA-code compared to the corresponding sRA and LDPC code.

\section{Conclusion}\label{concl}

We have designed in this paper new classes of structured LDPC codes with high code rate and low-complexity encoding. Based on specific classes of BIBDs, we obtained several infinite classes of $(k,r)$-regular LDPC codes with values of $k$ varying from 3 to 8 as well as an infinite class of $(k,r)$-regular LDPC codes for $k$ any prime power, all admitting flexible choices of $r$ and the code length. 
We have furthermore addressed sRA codes, proposing a generalized accumulator structure for higher column weights that replaces the conventional accumulator and leads to an encoding scheme, we have termed weight-$q$ RA code. This allowed an improved decoding performance closer to those of regular LDPC codes, along with the low encoding complexity of turbo-like codes. Compared to sRA codes, the relatively high error floors could be lowered significantly. The encoding scheme is applicable to our new construction techniques and may be used for further combinatorial constructions that lack an efficient encoding. The presented constructions arise from cyclic and resolvable BIBDs and allows to use an arbitrary number of block orbits or resolution classes for more flexibility in the code design. Therefore, we can adjust the rate and length of the codes independently and thus produce a wider range of codes. The proposed novel LDPC and systematic RA codes in this paper offer good structural properties, perform very well with the sum-product algorithm, and are suitable for applications, e.g., in high-speed applications in magnetic recording and optical communications channels.

\appendices

\section{Proof of Theorem~\ref{thm_CRCBIBD-LDPC Codes}}\label{APP_thm_CRCBIBD-LDPC Codes}

\begin{IEEEproof}
Let $p$ be a prime. We first assume that $k$ is odd. Then, by a result of Genma, Mishima and Jimbo~\cite{Gen97}, a $\mbox{CRCBIBD}(pk,k,1)$ exists, whenever there is an
$\mbox{RDF}(p,k,1)$. The following infinite ((i)-(iii)) and finite ((iv)) families of radical difference families are known (cf.~\cite{Bur95radical} and the references therein;~\cite{crc06}):

\begin{enumerate}

\item[(i)] An $\mbox{RDF}(p,3,1)$ exists for all primes $p \equiv 1$ (mod $6$).

\item[(ii)] Let $p=20t+1$ be a prime, let $2^e$ be the largest power of $2$ dividing $t$ and let $\varepsilon$ be a $5$-th primitive root of unity in $\mathbb{Z}_p$. Then an $\mbox{RDF}(p,5,1)$ exists if and only if $\varepsilon +1$ is not a $2^{e+1}$-th power in $\mathbb{Z}_p$, or equivalently $(11+5\sqrt{5})/2$ is not a $2^{e+1}$-th power in $\mathbb{Z}_p$.

\item[(iii)] Let $p=42t+1$ be a prime and let $\varepsilon$ be a $7$-th primitive root of unity in $\mathbb{Z}_p$. Then an $\mbox{RDF}(p,7,1)$ exists if and only if there exists an integer $f$ such that $3^f$ divides $t$ and $\varepsilon +1, \varepsilon^2 + \varepsilon +1,\frac{\varepsilon^2 + \varepsilon +1}{\varepsilon +1}$ are $3^{f}$-th powers but not $3^{f+1}$-th powers in $\mathbb{Z}_p$.

\item[(iv)] An $\mbox{RDF}(p,9,1)$ exists for all primes $p<10^4$ displayed in Table~\ref{table_RDF}.

\end{enumerate}

Moreover, in~\cite{Gen97} a recursive construction is given that implies the existence of a $\mbox{CRCBIBD}(\ell k,k,1)$ whenever $\ell$ is a product of primes of the form $p \equiv 1$ (mod $k(k-1)$).
In addition, a $\mbox{CRCBIBD}(5p,5,1)$ has been shown~\cite{Bur97resolvable} to exist for any prime $p \equiv 1$ (mod $20$) $< 10^3$. 

We now consider the case when $k$ is even:
In~\cite{Lam99}, a $\mbox{CRCBIBD}(4p,4,1)$ is constructed for any prime $p=20t+1$, where $t$ is an odd positive integer.
Furthermore, via the above recursive construction, a $\mbox{CRCBIBD}(4\ell,4,1)$ exists whenever $\ell$ is a product of primes of the form $p=12t +1$ and $t$ is odd. The result now follows.
\end{IEEEproof}

\section{Search Problem}\label{DiscreteLog}

Let $B_i=\{\omega^i, \omega^{i+2t}, \omega^{i+4t}\}$ be the $i$-th block of the difference family  \mbox{$F:=\{\omega^i G^{2t} \bmod p \ |\ 1\leq i \leq t\}$}
over $GF(p)$ with prime order $p=6t+1$. From the construction arises the following matrix
\[
\Delta F =
\left[
\begin{array}{cccc}
\omega^c&\omega^{c+1}&\cdots&\omega^{c+t-1}\\
\omega^{c+t}&\omega^{c+t+1}&\cdots&\omega^{c+2t-1}\\
\vdots&\vdots&&\vdots\\
\omega^{c+5t}&\omega^{c+5t+1}&\cdots&\omega^{c+6t-1}\\
\end{array}
\right]
\]
where the $i$-th column consists of the differences $\Delta B_i$, and  $\omega^c\equiv\omega(\omega^{2t}-1) \pmod p$. Note, that such an element $c$ must exist, since $\omega$ is a primitive element of $GF(p)$. Let $s\in \{0,...,t-1\}$ denote the index of the column that contains entry 1. The index $s$ indicates the position of the desired block $B_s$ containing the difference 1. Now, the following problems are equivalent:

\begin{enumerate}
	\item Computation of the discrete logarithm $c=Log_\omega(\omega^c)$
	\item Computation of $s$, such that $1\in \Delta B_s$.
\end{enumerate}

\begin{proof}
First, assume that the discrete logarithm is solved and thus $c$ is known. Then, $s=t-c\pmod{t}$.
Conversely, assume that $s$ is known. Let $r\in \{0,...,5\}$ be the row index of the 1-entry in $\Delta F$, such that $\Delta F_{(r,s)}=1$. Since there are only six possible rows, we can find $r$ in constant time. The solution of the discrete logarithm is then given by $c=6t-rt-s$.
\end{proof}

\section*{Acknowledgment}

The authors thank the anonymous referees for their careful reading and valuable insights that helped improving the presentation of the paper.

\begin{biography}[{\includegraphics[width=1in, height=1.25in,clip,keepaspectratio]{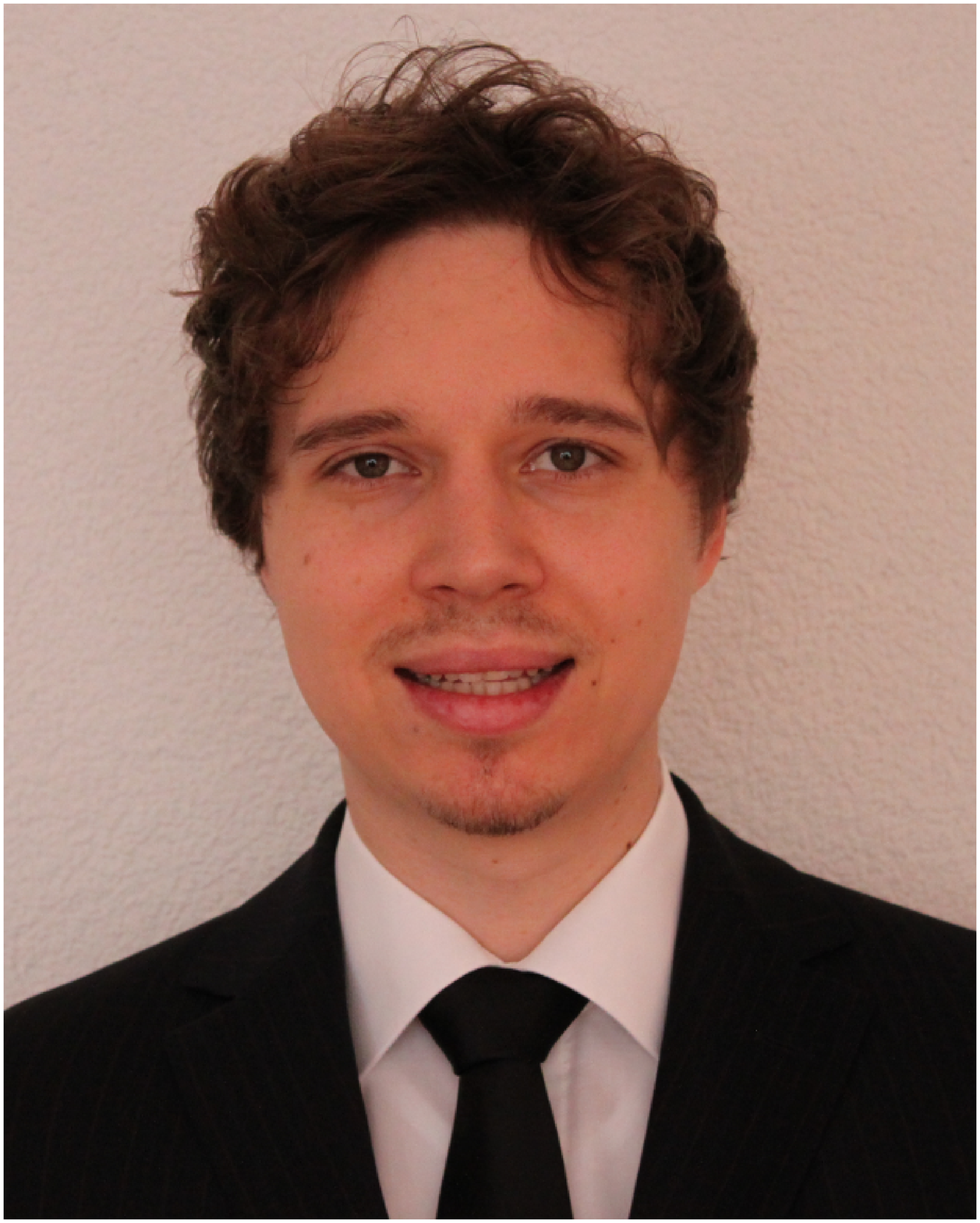}}]{Alexander Gruner} is a Ph.D. student in computer science at the Wilhelm Schickard Institute for Computer Science, University of T{\"u}bingen, Germany, where he is part of an interdisciplinary research training group in computer science and mathematics. He received the Diploma degree in computer science from the University of T{\"u}bingen in 2011. His research interests are in the field of coding and information theory with special emphasis on turbo-like codes, codes on graphs and iterative decoding.
\end{biography}

\vfill

\begin{biography}[{\includegraphics[width=1in, height=1.25in,clip,keepaspectratio]{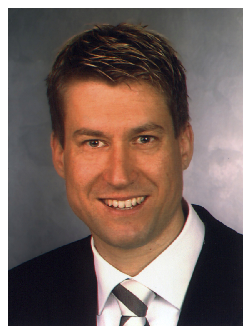}}]{Michael Huber} (M.'09) is a Heisenberg Research Fellow of the German Research Foundation (DFG) at the Wilhelm Schickard Institute for Computer Science, University of T{\"u}bingen, Germany, since 2008. Before that he had a one-year visiting Full Professorship at Berlin Technical University. 
He obtained the Diploma, Ph.D. and Habilitation degrees in mathematics from the University of T{\"u}bingen in 1999, 2001 and 2006, respectively. He was awarded the 2008 Heinz Maier Leibnitz Prize by the DFG and the German Ministry of Education and Research (BMBF). He became a Fellow of the Institute of Combinatorics and Its Applications (ICA), Winnipeg, Canada, in 2009.

Dr. Huber's research interests are in the areas of coding and information theory, cryptography and information security, combinatorics, theory of algorithms, and bioinformatics. Among his publications in these areas are two books, \emph{Flag-transitive Steiner Designs} (Birkh{\"a}user Verlag, Frontiers in Mathematics, 2009) and \emph{Combinatorial Designs for Authentication and Secrecy Codes} (NOW Publishers, Foundations and Trends in Communications and Information Theory, 2010). He is a Co-Investigator of an interdisciplinary research training group in computer science and mathematics at the University of T{\"u}bingen. 
\end{biography}

\enlargethispage{-5in}

\end{document}